\documentclass[%
 aip,
cp,  
 amsmath,amssymb,
 reprint,%
]{revtex4-2}

\usepackage{graphicx}
\usepackage{dcolumn}
\usepackage{bm}

\usepackage[utf8]{inputenc}
\usepackage[T1]{fontenc}
\usepackage{mathptmx} 
\usepackage{subcaption}
\usepackage{caption}

\begin{document}

\title{Development of Impedance Sheath Boundary Condition in Stix Finite Element RF Code}

\author{Christina Migliore} 
 \email[Corresponding author: ]{migliore@mit.edu}
 \affiliation{
  Massachusetts Institute of Technology, 77 Massachusetts Ave, Cambridge, MA, 02139 USA.
}
\author{Mark Stowell}%
 \email{stowell1@llnl.gov}
 \affiliation{
  Lawrence Livermore National Laboratory, 7000 East Ave, Livermore, CA 94550 USA.
}
\author{John Wright} 
 \email{jwright@psfc.mit.edu}
 \affiliation{
  Massachusetts Institute of Technology, 77 Massachusetts Ave, Cambridge, MA, 02139 USA.
}
\author{Paul Bonoli} 
 \email{bonoli@psfc.mit.edu}
 \affiliation{
  Massachusetts Institute of Technology, 77 Massachusetts Ave, Cambridge, MA, 02139 USA.
}

\date{\today} 

\begin{abstract}
Ion cyclotron radio frequency range (ICRF) power plays an important role in heating and current drive in fusion devices. However, experiments show that in the ICRF regime there is a formation of a radio frequency (RF) sheath at the material and antenna boundaries that influences sputtering and power dissipation. Given the size of the sheath relative to the scale of the device, it can be approximated as a boundary condition (BC). Electromagnetic field solvers in the ICRF regime typically treat material boundaries as perfectly conducting, thus ignoring the effect of the RF sheath. Here we describe progress on implementing a model for the RF sheath based on a finite impedance sheath BC formulated by J. Myra and D. A. D’Ippolito, Physics of Plasmas \textbf{22} (2015) which provides a representation of the RF rectified sheath including capacitive and resistive effects. This research will discuss the results from the development of a parallelized cold-plasma wave equation solver Stix that implements this non-linear sheath impedance BC through the method of finite elements in pseudo-1D and pseudo-2D using the MFEM library. The verification and comparison of the sheath BC from Stix with results from H. Kohno and J. Myra, Computer Physics Communications \textbf{220}, 129–142 (2017) will also be discussed.
\end{abstract}

\maketitle

\section{\label{sec:level1}Introduction}

As ion cyclotron radio frequency range (ICRF) heating becomes increasing used in fusion devices, the urgency of predicting and mitigating impurity generation that arises from it becomes increasingly important. In the ICRF regime, rectified RF sheaths are known to form at antenna and material edges influencing negative effects such as sputtering, a decrease in heating efficiency, and hot spots \cite{exp2, exp1}. With the goal of using steady state or long-pulsed fusion devices these adverse sheath effects will become increasingly important to predict and numerically model.

An effective way to simulate RF wave phenomenon in the scrape-off later of the tokamak is by using cold-plasma full-wave solvers. There exist many codes like these however a common issue is that the boundary conditions for the material surfaces use overly simplified models. Specifically, many codes use a conducting wall boundary condition that does not include the effects of the RF sheath rectification. In order to see how RF sheaths influence adverse effects in tokamaks in the ICRF regime, there is a need to include the physics of the RF sheath through the means of a BC in the edge solvers.

This paper introduces a newly developed parallized finite-element cold plasma RF solver mini-app called ``Stix'' aimed at bridging the gap of between modeling the global propagation behavior of the RF wave and the micro-scale physics of the RF sheath. The goal of this solver was to create a light-weight, but robust, mini-app that has easily adjustable boundary and plasma parameters such as density and temperature profiles, while including the incorporation of a finite impedance RF sheath BC formulated by J. Myra 2015 \cite{myra1}. With the addition of the RF sheath BC into Stix, this paper will discuss two examples comparing the results from this BC using previous literature, specifically Kohno et al. 2017 \cite{kohno17}.

\section{\label{sec:level1} COMPUTATIONAL MODEL}

\subsection{\label{sec:level2}Wave and RF sheath Physics}

An electromagnetic (EM) waves traveling in a plasma is given by the plasma wave equation as
\begin{equation}\label{waveequation}
\nabla \times \bar{\bar{\varepsilon}}^{-1}(\nabla \times \vec{H})-\omega^{2} \mu_{0} \vec{H}=\nabla \times \bar{\bar{\varepsilon}}^{-1} \vec{J}_{\text {ext }}
\end{equation}
where $\vec{J}_{\text {ext }}$ is the external current driven by the antenna. The plasma response to the EM wave is described by the plasma dielectric tensor, $\bar{\bar{\varepsilon}}$, given by
\begin{equation}
    \bar{\bar{\varepsilon}}=(\bar{\bar{I}}-\hat{b} \hat{b}) S+\hat{b} \hat{b} P+(i \hat{b} \times \bar{\bar{I}}) D
\end{equation}
where $\bar{\bar{I}}$ is the identity matrix and $\hat{b}$ is the unit vector of the background magnetic field defined as $\hat{b} = \vec{B}_{0}/|\vec{B}_{0}|$. Using Stix notation \cite{stix}, the coefficients of S, P, and D are defined as
\begin{equation}
S=1-\sum_{j} \frac{\omega_{p j}^{2}}{\omega^{2}-\Omega_{j}^{2}}, P=1-\sum_{j} \frac{\omega_{p j}^{2}}{\omega^{2}}, D=\sum_{j} \frac{\Omega_{j}}{\omega} \frac{\omega_{p j}^{2}}{\omega^{2}-\Omega_{j}^{2}}
\end{equation}

Given the thickness of the sheath relative to the wavelength of the wave, the RF sheath can be approximated as a boundary condition. This RF sheath BC is advantageous to encompass the physics of the plasma-wall interactions while maintaining a global RF wave code. A comprehensive representation of the RF sheath was formulated by J. Myra 2015 \cite{myra1} that includes both the resistive and capacitive components using a finite complex impedance is the tangential gradient of the RF sheath potential, $\phi$:

\begin{equation}
    E_{t} = \nabla_{t}(\phi) = \nabla_{t}\left ( -i \omega D_{n}z_{sh} \right )
\end{equation}

Here $z_{sh}$ is the sheath impedance which itself is a function of $\phi$, making the expression non-linear. This finite impedance encapsulates the micro-scale physics of the RF sheath that isn’t represented in a plasma wave equation solve. Given that there is no good analytic expression for $z_{sh}$, a parameterization code was created by J. Myra 2017 to estimate the $z_{sh}$ depending on the physics regime that could lean more capacitive or resistive based on the plasma physics parameters and the RF sheath potential, $\phi$, as an input \cite{myra2}.

\subsection{\label{sec:level3} Underlying Numerical Method of Stix}

Stix is a parallelized cold-plasma frequency domain RF wave solver that is built off of a finite element library called MFEM \cite{mfem} that solves Eq.\ \ref{waveequation}. Although Stix is inherently 3D due to MFEM, it is operational in pseudo-1D (Stix1D) and pseudo-2D (Stix2D) meaning that in the extra dimensions a phase factor is used. Stix’s lightweight nature was designed to make the implementation of various density, temperature, magnetic field, and collisional profiles easy.

For the finite-element aspect of the code, given that the plasma wave equation has inherently the curl operator within it, a natural choice is to use N\'{e}d\'{e}lec, also known as H(\textit{curl}), basis functions which are defined to have tangential continuity across element faces. Using these H(\textit{curl}) basis functions, here taken to be $\vec{W}_{i}$ and $\vec{W}_{j}$, Eq.\ \ref{waveequation} can be written in the weak form as

\begin{equation}\label{weakform}
\begin{split}
    \sum_{j} \{ \int_{\Omega}\left(\nabla \times \vec{W}_{i}\right) \cdot\left(\bar{\bar{\varepsilon}}^{-1} \nabla \times \vec{W}_{j}\right) d \Omega-\omega^{2} \mu_{0} \int_{\Omega} \vec{W}_{i} \cdot \vec{W}_{j} d \Omega + &\int_{\partial \Omega} \vec{W}_{i} \cdot\left(\hat{n} \times \bar{\bar{\varepsilon}}^{-1} \nabla \times \vec{W}_{j}\right) d \Gamma \} H_{j} \\
    &=-i \omega \int_{\Omega} \vec{W}_{i} \cdot\left(\nabla \times \bar{\bar{\varepsilon}}^{-1} \vec{J}_{\mathrm{ext}}\right) d \Omega
    \end{split}
\end{equation}
where the third term on the left-hand side denotes the boundary surface. Rewriting the boundary term in Eq.\ \ref{weakform} using Ampere’s Law gives the tangential electric field on the boundary, E$_{t}$. With a representation of E$_{t}$ in the boundary term, the RF sheath BC shown in Eq.\ \ref{sheathBC}, is able to be directly incorporated into the plasma wave equation.

The RF sheath BC requires the normal component of the electric displacement, $D_{n}$, to be known before the solve. One of the advantages to solving for the magnetic flux field, $\vec{H}$, rather than the electric field, $\vec{E}$, is that the RF sheath BC can be rewritten in terms of $\vec{H}$ as shown in Eq.\ \ref{sheathBC} allowing both Eq.\ \ref{weakform} and Eq.\ \ref{sheathBC} to be solved in one step. 

\begin{equation}\label{sheathBC}
    \phi=-i \omega D_{n} z_{s h}(\phi)=-i \omega \hat{n} \cdot\left(\frac{\vec{\nabla} \times \vec{H}}{-i \omega}\right) z_{s h}=\hat{n} \cdot(\vec{\nabla} \times \vec{H}) z_{s h}
\end{equation}


Stix solves Eq.\ \ref{weakform} and Eq.\ \ref{sheathBC} simultaneously for $\vec{H}$ and $\phi$ using a block matrix solve. For the wave equation block, SuperLU calculates the Schur complement which then is used to iteratively solve the entire system matrix using GMRES. Once the $\vec{H}$ and the corresponding RF sheath potential, $\phi$, are found, $\phi$ is passed into the parameterization code given by J. Myra 2017 \cite{myra2} that in return gives a new sheath impedance value, $z_{sh}$, which then updates the boundary term in Eq.\ \ref{sheathBC}. With the new updated RF sheath potential, the whole $\vec{H}$ field is solved all over again on the entire domain. The code does a fixed point iteration on $\phi$ until the desired convergence criterion of the errors between solves is met, usually taken to be $|\phi_{n+1} - \phi_{n}| = 10^{-5}$.

\begin{figure}[h]
\includegraphics[width=1.0\columnwidth]{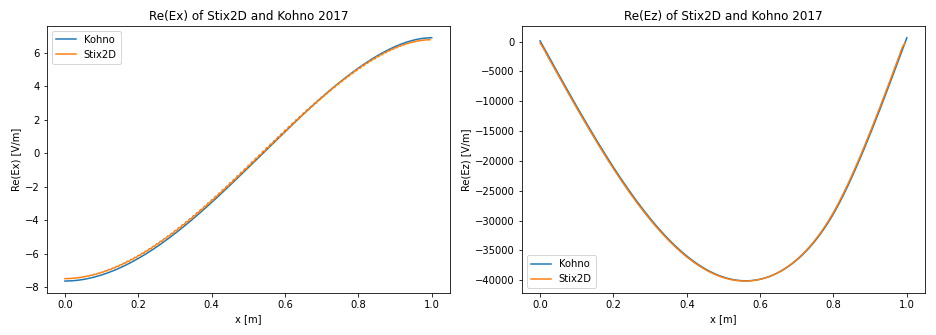}
\caption{\label{fig:fig_3} Comparison of the real E$_{x}$ and E$_{z}$ fields between Stix2D (\textbf{orange line}) and Kohno et al. 2017 (\textbf{blue line}) \cite{kohno17}.}
\end{figure}

\begin{table}[h]
\caption{\label{tab:table1} Comparison of the magnitude of the RF sheath potential, $|V_{RF}|$, in the 1D Kohno et al 2017 \cite{kohno17} case against Stix's results.  }
\begin{ruledtabular}
\begin{tabular}{c c c}
Location of BC &  |$V_{RF}$| of Stix &  |$V_{RF}$| of Kohno et al. 2017 \\
\hline
Left-hand side & 24.86 V & 25.41 V \\
Right-hand side & 99.1 V & 101.80 V \\
\end{tabular}
\end{ruledtabular}
\end{table}

\section{\label{sec:level1}Results}

With Eq.\ \ref{sheathBC} incorporated into the Stix solve, verification that the resulting sheath potentials give the expected result needed to be done. Two example cases for verification were compared to Stix’ results which were both taken from Kohno et al. 2017 \cite{kohno17}. 

First case was a 1D problem taken to have a domain in x extending from 0 to 1 m with an antenna placed at 0.8 m. The parameters were taken to be a constant density of $3 \times 10^{17} m^{-3}$, $k_{z} = 10.8 \ m^{-1}$, a constant background magnetic field of $B_{0} = 1 \hat{x}$ T, antenna amplitude of 560 A/m, and 80 MHz. On both ends of the domain the RF sheath BC was placed. Fig. \ref{fig:fig_3} shows the resulting comparison of both the real E$_{x}$ and E$_{z}$ components of the electric field solutions from Stix1D and Kohno et al.\ 2017 \cite{kohno17}. Both electric fields show good agreement with one another. Additionally, the comparison of the magnitudes of the RF sheath potentials were shown to agree with one another within a 2 V difference as shown in Table \ref{tab:table1}. 

The second case was a rectangular domain 2D problem that extended from 0 to 1.2 m in the x-direction and 0 to 0.2 m in the y-direction with an antenna set at 1 m with a height of 0.05 m in the y-direction. Conducting wall BCs were taken on all sides expect for the right-most boundary behind the antenna in which the RF sheath BC was applied. For this case, it was a slow wave propagating directly into the wall with the parameters taken to be a constant density of $n_{e} = 10^{17} \ m^{-3}$, $k_{z} = 320 \ m^{-1}$, a constant background magnetic field of $B_{0} = 1 \hat{x}$ T, and 80 MHz. In addition, the antenna was taken to have a cosine squared profile with an amplitude of K$_{max}$ = 10 A/m pointing in the $\hat{y}$ and an artificial collisional frequency profile imposed as $\nu(x) = 3 \times 10^{11}\exp^{-x/0.1}$ s$^{-1}$ in order to have no reflections from the left-hand side.
\begin{figure}[h]
     \centering
     \begin{subfigure}[b]{0.59\textwidth}
         \centering
         \includegraphics[width=\textwidth]{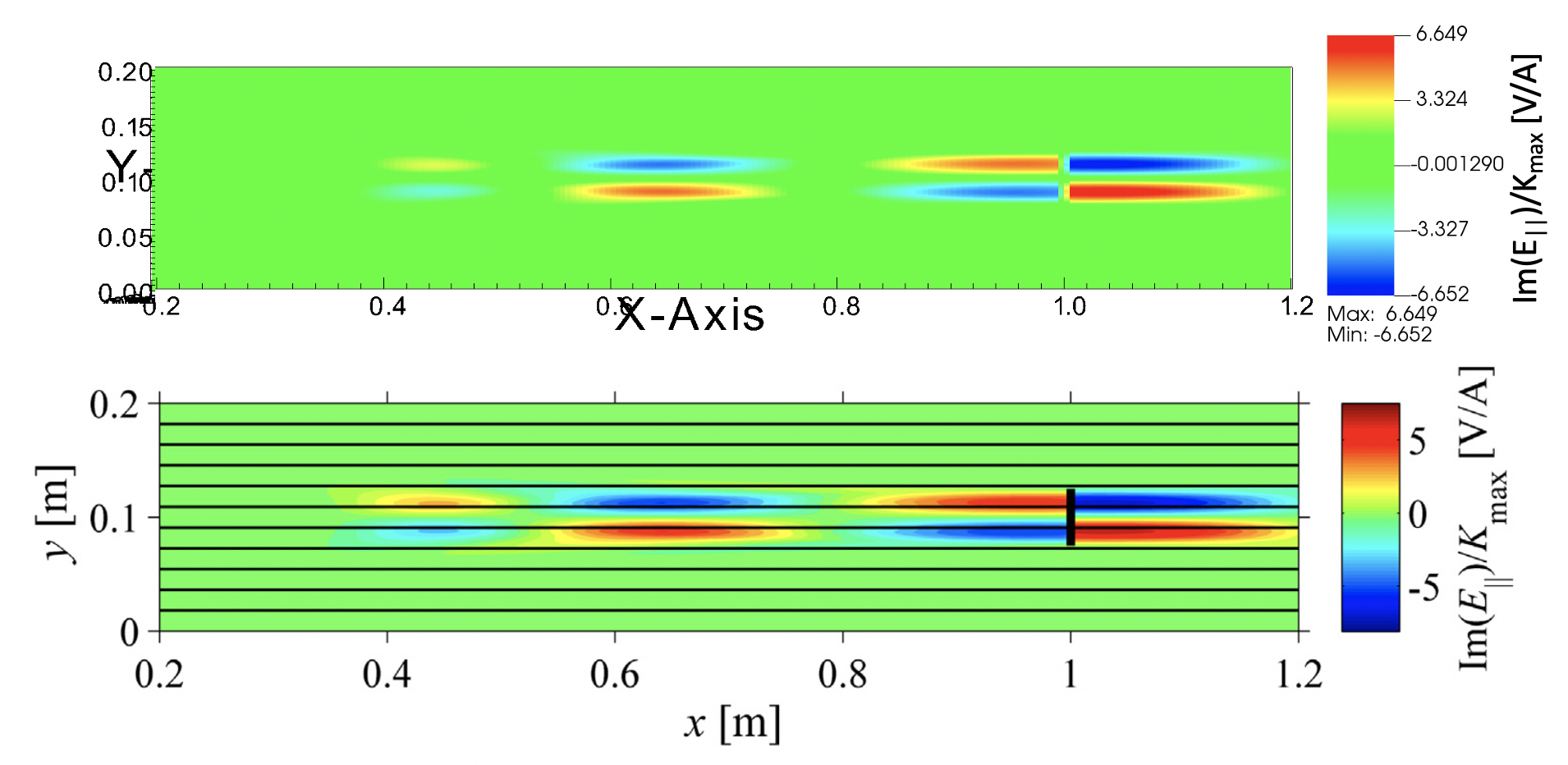}
         \caption{Comparison of the 2D color plot of Im(E$_{||}$/K$_{max}$) of Stix2D on the top and Kohno et al. 2017 on the bottom \cite{kohno17}.}
         \label{fig:y equals x}
     \end{subfigure}
     \hfill
     \begin{subfigure}[b]{0.4\textwidth}
         \centering
         \includegraphics[width=\textwidth]{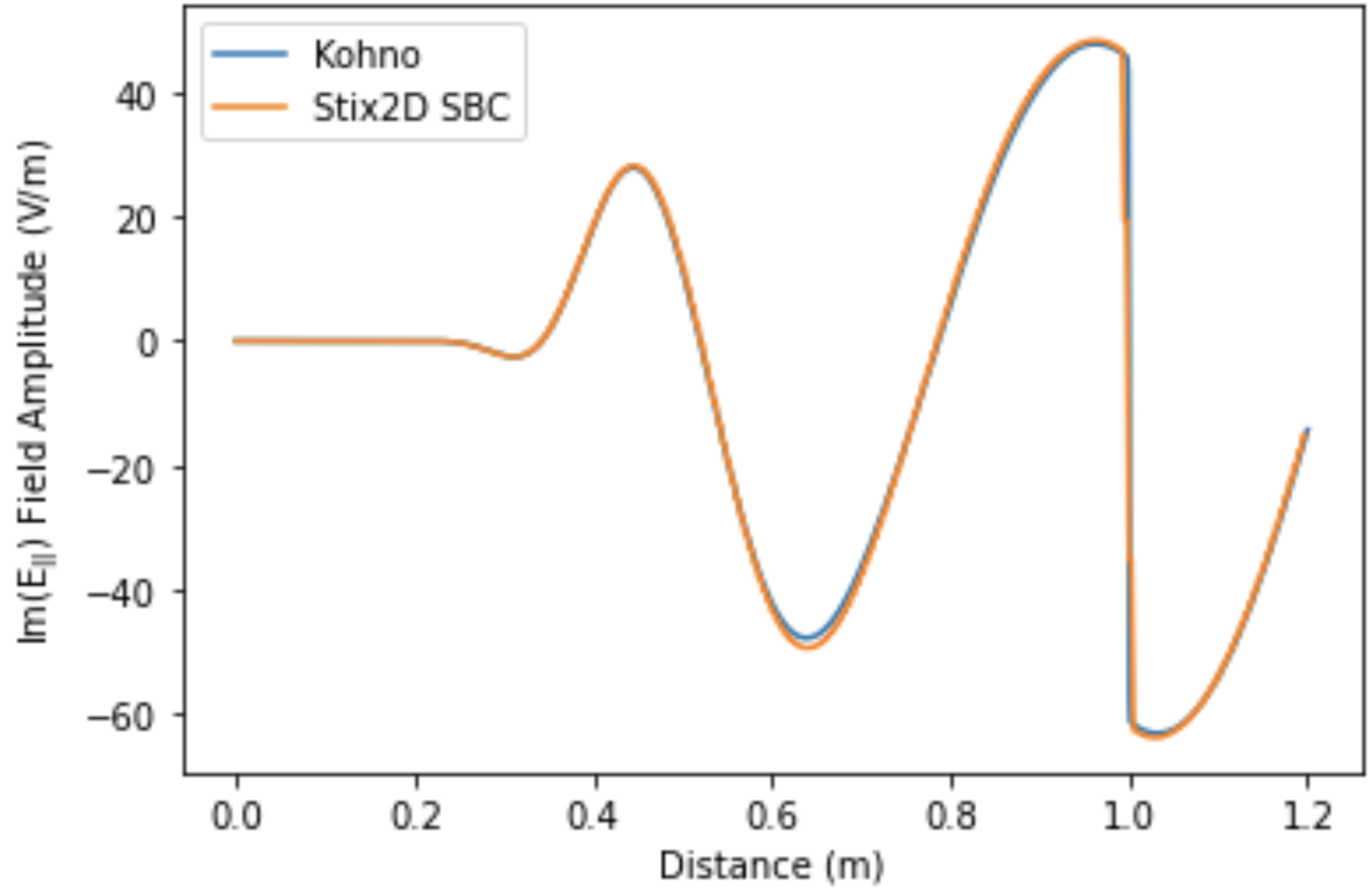}
         \caption{Comparison of a lineout taken at y=0.11 between Stix2D (\textbf{orange line}) and Kohno et al. 2017 (\textbf{blue line}) \cite{kohno17}.}
         \label{fig:three sin x}
     \end{subfigure}
     \caption{\label{fig:fig_6} Comparison of the imaginary parallel electric field, Im$(E_{||})$, between Kohno et al. 2017 Fig. 6 and Stix2D as a 2D pseudo-color plot (\textbf{left}) and a lineout along the x-direction (\textbf{right})\cite{kohno17}.}
\end{figure}

Fig.\ \ref{fig:fig_6}a shows the resulting 2D solution of the imaginary parallel electric field of both Stix2D and Kohno et al. 2017. A closer look at the fields using a lineout along y = 0.11, Fig.\ \ref{fig:fig_6}b, shows agreement between both solutions including the region at the sheath boundary surface on the right-hand side of the domain.

\begin{figure}
\includegraphics[width=0.7\columnwidth]{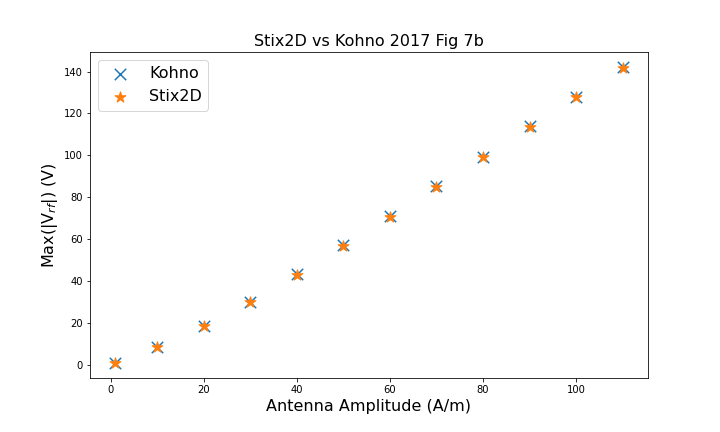}
\caption{\label{fig:fig_7b} Comparison of the maximum magnitude of the RF sheath potential, $|V_{RF}|_{max}$, along the right-most boundary behind the antenna between Stix2D and Kohno et al. 2017 \cite{kohno17} for various antenna amplitudes.}
\end{figure}

For the 2D example case just discussed, Kohno et al. 2017 additionally sweeps various antenna amplitudes starting from 0 A/m to 140 A/m to see how the RF sheath potential behavior changes. Shown in Fig.\ \ref{fig:fig_7b} is the comparison of Kohno's results and Stix2D's results from the same antenna amplitude sweep. This plot shows that the results found from Stix are consistent with the results from Kohno et al.\ 2017 \cite{kohno17} for different regimes of the RF sheath that are dependent on the strength of the antenna current.

\section{\label{sec:level1}Conclusion}

In this paper, a cold-plasma finite-element RF solver that includes a finite impedance RF sheath boundary condition developed by J. Myra 2015 \cite{myra1}, ``Stix,'' is introduced. Adding this RF sheath BC involved solving for the magnetic field, $\vec{H}$, simultaneously in the plasma wave equation and the sheath potential equation. Using Kohno et al. 2017 examples cases, verification of this BC implementation was conducted and found to have good agreement. Moving forward, future work involving Stix will shift towards simulating more realistic geometric scenarios that solve for the sheath potential looking at both near and far-field sheaths.

Although it is important to know the voltages on material surfaces due to rectification, an important problem facing ICRF heating is the enhancement of impurities generated at the plasma-material boundary due to the rectified sheaths \cite{danmyra1}. Having an integrated ICRF full-wave solver that calculates sputtering yields and rectified potentials will allow for in depth and realistic investigations into where the largest sputtering occurs and the differences in wall materials. One of the long-term aims behind the creation of Stix is to have the ability to easily couple the RF sheath potential solve to an impurity generation code, such as RustBCA \cite{rust}, in order to have impurity fluxes calculated based on the DC rectified potential.

\begin{acknowledgments}
This work was supported by the U.S. Department of Energy Scientific Discovery through Advanced Computing Initiative, Contract Number DE-SC0018090.
\end{acknowledgments}

\nocite{*}

\end{document}